# ENERGY SPECTRUM OF THE DIRAC OSCILLATOR IN THE COULOMB FIELD

**D.A. Kulikov, R.S. Tutik**

Dniepropetrovsk National University, Physics Department,
13 Naukova Street, 49050, Dniepropetrovsk, Ukraine
e-mail: kulikov_d_a@yahoo.com

*Abstract* – **The spectrum of the Dirac oscillator perturbed by the Coulomb potential is considered. The Regge trajectories for its bound states are obtained with the method of ℏ-expansion. It is shown that the split of the degenerate energy levels of the Dirac oscillator in the Coulomb field is approximately linear in the coupling constant.**

## I. INTRODUCTION

The current interest in the Dirac oscillator is motivated by its possible applications in nuclear and hadron physics (see [1] and references therein). This model is a version of harmonic oscillator for relativistic spin ½ particles. It has the exact solution and the degenerate spectrum of eigenenergies $E^2 - m^2 = \hbar m \omega (4n + 4l + 2)$ [2]. However, the unphysical degeneracy of the spectrum is removed in the perturbative external field. The goal of the present work is to consider the split of the degenerate energy levels of the Dirac oscillator caused by the external Coulomb field.

## II. THE METHOD

In the Dirac equation, we arrive at the Dirac oscillator with the non-minimal substitution $\vec{p} \to \vec{p} - im\omega\beta\vec{r}$ where $\omega$ is the oscillator frequency and $\beta$ is the usual Dirac matrix. Then the model of the Dirac oscillator perturbed by the Coulomb potential $V(\vec{r}) = -b/r$ is described by the equation (in units with $c = 1$)

$$[\vec{\alpha} \cdot (\vec{p} - im\omega\beta\vec{r}) + m\beta - \frac{b}{r}]\Psi = E\Psi. \qquad (1)$$

After separating the variables, this equation reduces to the system

$$\hbar G'(r) + (\frac{\hbar \chi}{r} + m\omega r)G(r) - (E + m + \frac{b}{r})F(r) = 0,$$

$$\hbar F'(r) - (\frac{\hbar \chi}{r} + m\omega r)F(r) + (E - m + \frac{b}{r})G(r) = 0, \qquad (2)$$

where $\chi = s(j + 1/2)$, with $j = l - s/2$ and $s = \pm 1$, is the standard invariant [3].

In order to solve the bound state problem defined by equations (2), we extend the method of the ℏ-expansion for the Regge trajectories [4],[5] to the case of the Dirac oscillator interaction.

Let us recall that within the framework of the non-relativistic quantum mechanics, the Regge trajectory is defined as the dependence of the orbital quantum number on the energy, $\alpha(E) = \hbar l(E)$, at the fixed value of the radial quantum number $n$. For the Dirac equation, the role of the Regge trajectory plays the invariant quantity $\alpha(E) = \hbar \chi(E)$. Since the states of the Dirac oscillator with $\chi = -(j + 1/2)$ are unphysical, we will consider only the case of $\chi = +(j + 1/2) = l$.

Following the work [5], we transform equations (2) into the non-linear Riccati equation with respect to the logarithmic derivative $C(r) = \hbar G'(r)/G(r)$ that permits us to derive the recursion relations. We seek the solution of the Riccati equation as the series expansions in powers of the Planck's constant $\hbar$

$$C(r) = \sum_{k=0}^{\infty} C_k(r)\hbar^k, \quad \alpha(E) = \sum_{k=0}^{\infty} \alpha_k(E)\hbar^k. \qquad (3)$$

Because the wave function of the $n$-th radially excited state has exactly $n$ nodes on the real axis its logarithmic derivative should satisfy the quantization condition

$$\frac{1}{2\pi i}\oint C(r)dr = n\hbar, \tag{4}$$

where the contour of the integration encloses only the above mentioned nodes.

This condition must be supplemented with a rule of passing to the classical limit for the radial ($n$) and orbital ($l$) quantum numbers which are the specific quantum notions. We use the following rule

$$\hbar \to 0, \; n = const, \; l \to \infty, \; \hbar n \to 0, \; \hbar l = const \tag{5}$$

that allows us to rewrite the quantization condition (4) as

$$\frac{1}{2\pi i}\oint C_1(r)dr = n, \quad \frac{1}{2\pi i}\oint C_k(r)dr = 0, \; k \neq 1, \tag{6}$$

From the physical point of view, the rule (5) means that in the classical limit the particle is located in the bottom of the effective potential and, consequently, moves along the stable circular orbit. Then the classical orbital momentum of the particle, which is the first approximation to its Regge trajectory, takes the form

$$\alpha_0(E) = \sqrt{m^2\omega^2 r_0^4 + bEr_0 + b^2}, \tag{7}$$

where $r_0$ is the radius of the circular orbit.

For convenience of consideration, let us introduce the new variable $x = (r - r_0)/r_0$ that is the deviation from the minimum of the effective potential. According to the general scheme of the ℏ-expansion method [4],[5], unknown functions $C_k(r)$ can be represented by the Laurent series

$$C_k(r) = x^{1-2k}\sum_{i=0}^{\infty} C_i^k x^i. \tag{8}$$

Then from the Riccati equation we arrive at the recursion relation

$$C_k^i = \frac{1}{2C_0^0}\{\theta(i-2k+2)[\frac{(-1)^i(3-2k+i)}{r_0^2}(\alpha_{k-1}(E) + \sum_{j=0}^{k}\alpha_j(E)\alpha_{k-j}(E)) + m\omega(y_i + y_{i-1})\delta_{k,1}$$
$$+ \frac{\alpha_{k-1}(E)}{r_0^2}\sum_{p=0}^{i-2k-2}(-1)^{i-p}y_p] - \frac{3-2k+i}{r_0}C_i^{k-1} - \sum_{j=1}^{k-1}\sum_{p=0}^{i}C_p^j C_{i-p}^{k-j} - 2\sum_{p=1}^{i}C_p^0 C_{i-p}^k + \frac{1}{r_0}\sum_{p=0}^{i}y_p C_{i-1-p}^{k-1}\}, \tag{9}$$

where $y_i = [-(m+E)/(m+E+b/r_0)]^{1+i} - (-1)^{1+i}$ and $\delta_{k,i}$ is the Kronecker delta.

Taking into account the quantization conditions (6), transformed by the theorem of residues into the form

$$C_0^1 = n/r_0, \; C_{2k-2}^k = 0, \; k \neq 1, \tag{10}$$

the final expression for the coefficients of the ℏ-expansions of Regge trajectories $\alpha_k(E)$ with $k \geq 1$ reads as

$$\alpha_k(E) = \frac{1}{2(\alpha_0(E) + m\omega r_0^2)}[r_0 C_{2k-2}^{k-1} + r_0^2\sum_{j=0}^{k}\sum_{p=0}^{2k-2}C_p^j C_{2k-2-p}^{k-j} - r_0\sum_{i=0}^{2k-3}y_i C_{2k-3-i}^{k-1}$$
$$+ \delta_{k,1}m\omega r_0^2(1-y_0) - \alpha_{k-1}(E)(1+y_0) - \sum_{j=1}^{k-1}\alpha_j(E)\alpha_{k-j}(E)]. \tag{11}$$

The first two corrections to the Regge trajectories computed with this recursion formula provide a sufficient accuracy for practical applications.

The parent ($n=0$) and the first two daughter ($n=1$, $n=2$) Regge trajectories calculated in relativistic units ($\hbar=c=1$) with the values of the parameters $m=1$, $\omega=10$, $b=0.5$ are shown in Fig. 1. The points on the Regge trajectories correspond to positions of the bound states designated by the standard spectroscopic manner.

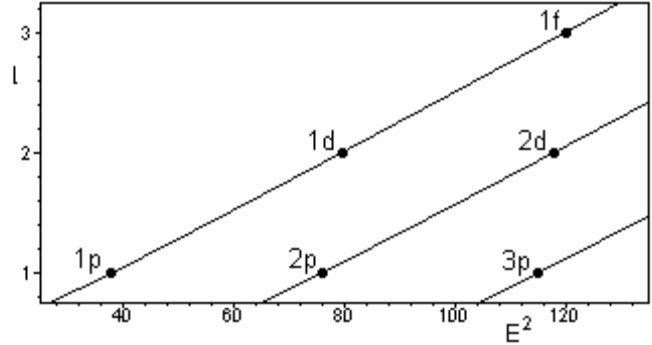

Fig. 1.   Regge trajectories of the bound states of the Dirac oscillator perturbed by the Coulomb potential.

From Fig. 1 we see that the Regge trajectories are approximately linear in the squared energy. It will be recalled that the exact linearity is realized when the Coulomb interaction is turned off.

The addition of the Coulomb potential splits the degenerate energy levels of the pure Dirac oscillator, and instead of $E_{1d}=E_{2p}$, $E_{1f}=E_{2d}=E_{3p}$ etc, we have now the inequalities $E_{1d}>E_{2p}$, $E_{1f}>E_{2d}>E_{3p}$.

It was found that magnitudes of the splits $E_{1d}-E_{2p}$, $E_{1f}-E_{3p}$, $E_{2d}-E_{3p}$ depend on the value of the Coulomb coupling constant $b$ and are nearly linear in $b$ at sufficiently small its values $b^2\omega/m<1$.

It should be pointed that there is a critical value of the Coulomb coupling constant: $b=1$. If $b>1$, the levels $1p, 2p, 3p,...$ disappear and we have the case of spontaneous pair creation.

### III. Conclusion

We have considered the bound-state problem for the Dirac oscillator perturbed by the Lorentz-vector Coulomb potential. To derive the Regge trajectories of the bound states of this system, the semiclassical technique of the asymptotical ℏ-expansions has been applied.

The dependence of the energy spectrum on the Coulomb coupling constant has been studied. It has been found that magnitudes of the splits of energy levels are nearly linear in the Coulomb coupling constant. For the case of the Lorentz-scalar Coulomb potential, the similar behaviour of the Regge trajectories has been observed.

### References

[1.]   R. Lisboa, M. Malheiro, A. S. de Castro, P. Alberto, M. Fiolhais, "Pseudospin symmetry and the relativistic harmonic oscillator," *Physical Reviews*, vol. C-69, 024319, February 2004.
[2.]   M. Moshinsky and A. Szczepaniak, "The Dirac oscillator," *Journal of Physics: Mathematical and General*, vol. A-22, pp. L817-819, September 1989.
[3.]   V. B. Berestetskii, E. M. Lifshitz, L. P. Pitaevskii, *Kvantovaya Elektrodinamika*, 2nd ed., Moskva, Nauka, 1980.
[4.]   S. S. Stepanov, R. S. Tutik, "The ℏ-expansion for the bound states of the Schrődinger equation," *Teor. i Matem. Fiz.*, vol. 90, pp. 208-217, February 1992..
[5.]   S. S. Stepanov, R. S. Tutik, "The ℏ-expansion for Regge trajectories: 2. Relativistic equations," preprint *ITP-92-14E,* Kiev, April 1992.